\documentclass[aps,twocolumn,nofootinbib,showpacs, prd]{revtex4}
\usepackage{graphicx}
\usepackage{amsfonts}
\usepackage{amssymb}
\usepackage{amsbsy}
\usepackage{amsmath}
\usepackage{latexsym}
\usepackage{natbib}
\usepackage{bm}
\usepackage{color}
\usepackage{float}

\begin{document}
\title{A Collapsing Distribution of Axion Dark Matter}

\author{Soumya Chakrabarti\footnote{soumya.chakrabarti@saha.ac.in}}
\affiliation{Theory Division,\\
Saha Institute of Nuclear Physics, \\
Kolkata 700064, West Bengal, India
}
 
\pacs{04.20.−q, 04.20.Dw, 04.20.Jb, 04.70.Bw}

\date{\today}

\begin{abstract}
The manuscript deals with an interacting scalar field that mimics the evolution of the so-called Axion Scalar Dark Matter or Axion like particles with ultra-light masses. It is discussed that such a scalar alongwith an ordinary fluid description can collapse under strong gravity. The end state of the collapse depends on how the Axion interacts with geometry and ordinary matter. For a self-interacting Axion and an Axion interacting with geometry the collapse may lead to a zero proper volume singularity or a bounce and total dispersal of the Axion. However, for an Axion interacting with the ordinary fluid description, there is no formation of singularity and the Axion field exhibits periodic behavior before radiating away to zero value. Usually this collapse and dispersal is accompanied by a violation of the Null Energy Condition.
\end{abstract}

\maketitle

\section{Introduction}
General Theory of Relativity (GR) describes gravity as an artifact of geometry and is full of counter-intuitive ideas and solutions. In particular, we focus on an idea that a massive stellar distribution written under the scope of GR can in principle collapse indefinitely to form a geodesic incompleteness, where curvature scalars diverge. This region is famously known as a Spacetime Singularity. An amendment is sometimes proposed in the theory in the form of a `Cosmic Censorship Conjecture (CCC)’ \cite{penrose}. The conjecture predicts the singularity to remain covered through a development of trapped surface, in process producing a Black Hole end-state \cite{os}. However, without any concrete proof of the conjecture, nothing stands on paper to favor the Black Holes over it's sibling outcome, the Naked Singularities, who can share information with a faraway observer \cite{collapse1}. Indeed, GR allows existence of such solutions where a collapsing star fails to form any trapped surface/horizon and keeps the ultimate singularity visible \cite{collapse2}. This is no doubt problematic for the predictability of events in the spacetime. The time evolution of a collapsing stellar body is the subject that inspires these inconclusive questions and the topic receives significant attention even today, in GR and it's viable modifications as well \cite{collapse3}. \\

Modified gravity is a prospective avenue to play with the idea of CCC as here one can try to produce non-trivial geometric corrections, affecting the nature and understanding of the conjecture. Most of the modified gravity models are also motivated from cosmological requirements. They can describe the observed cosmic acceleration of our universe. A particularly popular candidate in this regard is a time evolving interacting scalar field that can produce an effective repulsive effect and fit in nicely as the so-called Dark Energy (DE) component of the universe. Any such model must also explain the large scale structures and the proper profile of relic radiation from big bang for the universe. This requires an additonal feature which can drive the cosmic deceleration and describe the origin of the so-called Dark Matter (DM) component. The nature of this DM is one of the most studied cosmological question of modern era \cite{bertone}. A pressureless `cold dark matter' or the $\Lambda$CDM can explain the large scale structure formation of the universe \cite{planck}, however, numerical and observational evidences from galaxies and their clusters have imposed strong constraints on this description. Many alternative models work well to fit in as DM, for example the Weakly interacting massive particle models \cite{queiroz, hooper}, the warm DM \cite{collin, villa}, self interacting dark matter models \cite{loeb, spergel}). One particular approach has received recurring attention in the past decade, a Scalar Field Dark Matter (SFDM). This involves a time dependent scalar field endowed with an interaction potential that fits in the role of DM. It was first put forward Ji and Sin \cite{jisin} and Sin \cite{sin}, and received considerable treatment thereafter by Sahni and Wang \cite{sahni}, Hu et al. \cite{hu}, Matos et al. \cite{matos1}, Matos and Urena-Lopez \cite{matos2}, Arbey et al. \cite{arbey} and Matos and Arturo Ureña-López \cite{matos3}. \\

We are interested in a particular member of the SFDM family, the QCD inspired `Axion', described by a field \cite{siki} or a system of Axion-like particles (ALP) with ultra-light masses \cite{axiverse}. An Axion can be described by a scalar field with a very low mass and a self-interaction potential. We discuss that these fields or ALP-s can collapse due to strong gravity and form infinitely dense points in spacetime or singularities, gravitationally bound objects or dynamical equilibrium states \cite{cores, clusters} through a scalar field collapse. On it's own, collapse of scalar fields is an interesting subject which tries to answer if the CCC is satisfied by such fundamental matter fields \cite{scalarcollapse}. Moreover, it can be proved that an interacting scalar field can reproduce the evolution of different type of matter distributions \cite{moss1}. The subject gained additional popularity after the discovery of the `Critical Phenomena' in the collapse of zero mass scalar fields \cite{chop, brady, gund}. The phenomenon indicates that a scalar field can collapse and form either a singular end-state or disperse away to zero depending on a finite number of critical parameters. \\

We study the collapse of the dark matter candidate Axion or ALP-s. We write their description as an interacting scalar field. Such an Axion distribution has received some limited interest only recently where possible end states of the collapse is identified \cite{moss2} numerically. We work with a spatially homogeneous Oppenheimer-Snyder like geometry and discuss three examples with step by step generalization of the scalar interaction. (a) A minimally coupled Axion field with self-interaction potential. (b) An Axion field coupling non-minimally with geometry alongwith a self-interaction potential. (c) An Axion field interacting non-minimally with both geometry and ordinary matter. These considerations are motivated from a thought that during the final stages of a gravitational collapse, where the strength of gravity and the spacetime curvature is expected to be quite high, non-trivial interactions between scalar field and geometry and even normal matter can exist. A similar scalar configuration is popular from a cosmological purview, known as Chameleon Fields, who can provide a very smooth transition from a decelerated to an accelerated phase of expansion of the universe \cite{cham}. Apart from the Axion field and it's interactions, we also consider a perfect fluid distribution without any restriction over the equation of state parameter at the outset. This is motivated from a requirement to describe the distribution of DE by the aforementioned fluid distribution. No concrete information on the distribution of the DE component is known apart from the speculation that it does not cluster below Hubble scale. The main focus of the present work is the time evolution of collapsing Axion Dark Matter and formation of gravitationally bound end-states. However, the collapsing perfect fluid alongwith the scalar field motivates one to think about the possibility of clustering of DE.

\section{Theoretical Setup and Methodology}
We assume that the dark matter distribution is overall given by a scalar field with a self-interaction. We write the generic Axion potential $V(\phi)$ as \cite{moss2}
\begin{equation}\label{axionpot}
V(\phi) = m^{2}f^{2}\left[1 - cos\Big(\frac{\phi}{f}\Big)\right].
\end{equation}
$m$ is the mass and $f$ is the decay constant of the ALP-s. Usually the ALP dark matter model predicts the density parameter to be given by
\begin{equation}
\Omega_{ALP} \sim 0.1 \left(\frac{f}{10^{17}GeV}\right)^{2} \left(\frac{m}{10^{-22}eV}\right). 
\end{equation}
In the present work we take $m$ and $f$ as free parameters and assume $f >> m$. Under this assumption we write the Axion potential in Eq. (\ref{axionpot}) as
\begin{equation}\label{ourpot}
V(\phi) = \frac{1}{2}m^{2}\phi^{2} - \frac{m^2}{24f^2}\phi^4.
\end{equation}
This form has an interesting resemblance with the Higgs potential, which has played significant role in the cosmological context as in inflation \cite{moss3} and cosmological reconstruction of modified gravity \cite{sc1}. In the rest of the manuscript, we study scalar field collapse where the self-interaction of the scalar field is given by Eq. (\ref{ourpot}). Three different setups are considered, (a) a minimally coupled self-interacting Axion, (b) a non-minimally coupled self-interacting Axion and (c) a third case where the Axion interacts with the ordinary matter. For all three cases, we choose a spatially flat homogeneous metric of Oppenheimer-Snyder type, given by
\begin{equation}\label{metric}
ds^2 = dt^2 - a(t)^2(dr^2+r^2d\Omega^2).
\end{equation}

We restrict our study to collapsing models only. This means that radius of the two-sphere is a monotonically decreasing function of time, i.e., $\dot{a} < 0$. We follow an unconventional methodology compared to the standard methods of studying exact solutions of scalar field cosmology or collapse. We deal with the Klein Gordon equation of the Axion scalar field by incorporating a purely mathematical property of general second order differential equations with variable coefficients. The propery involves point transforming the equations into an integrable form and is derived from the symmetry analysis of a general classical anharmonic oscillator equation system \cite{duarte, euler1, euler2, harko}. The general equation is written as
\begin{equation}\label{anharmonic}
\ddot{\phi}+f_1(t)\dot{\phi}+ f_2(t)\phi+f_3(t)\phi^n=0.
\end{equation}
$f_1$, $f_2$ and $f_3$ are unknown functions of some variable, of t at this point. $n$ is a constant. A transformation of this equation into an integrable form requires a pair of point transformations and the condition $n\notin \left\{-3,-1,0,1\right\} $ to be satisfied. Moreover, the coefficients must satisfy the condition

\begin{eqnarray}\label{criterion1}\nonumber
&&\frac{1}{(n+3)}\frac{1}{f_{3}(t)}\frac{d^{2}f_{3}}{dt^{2}} - \frac{(n+4)}{\left( n+3\right) ^{2}}\left[ \frac{1}{f_{3}(t)}\frac{df_{3}}{dt}\right] ^{2} \\&&\nonumber
+ \frac{(n-1)}{\left( n+3\right) ^{2}}\left[ \frac{1}{f_{3}(t)}\frac{df_{3}}{dt}\right] f_{1}\left( t\right) + \frac{2}{(n+3)}\frac{df_{1}}{dt} \\&&
+\frac{2\left( n+1\right) }{\left( n+3\right) ^{2}}f_{1}^{2}\left( t\right)=f_{2}(t). 
\end{eqnarray} 

The point transformations are written as 
\begin{eqnarray}\label{criterion2}
\Phi\left( T\right) &=&C\phi\left( t\right) f_{3}^{\frac{1}{n+3}}\left( t\right)
e^{\frac{2}{n+3}\int^{t}f_{1}\left( x \right) dx },\\
T\left( \phi,t\right) &=&C^{\frac{1-n}{2}}\int^{t}f_{3}^{\frac{2}{n+3}}\left(
\xi \right) e^{\left( \frac{1-n}{n+3}\right) \int^{\xi }f_{1}\left( x
\right) dx }d\xi ,\nonumber\\
\end{eqnarray}%
where $C$ is a constant. Using this property, we solve the Axion scalar evolution equation assuming it's integrability at the outset and use the other field equations to assess the constraints on the fluid energy density components. The main motivation of assuming this integrability comes out of a pure mathematical curiosity. However, by no means this produces unphysical solutions. The scope of this approach has been discussed at length quite recently, in the context of simple scalar field collapse \cite{sc2}, scalar-gauss-bonnet gravity \cite{sc3} and cosmological reconstruction of modified theories of gravity \cite{sc1}. 
                      
\section{Minimally Coupled Axion}
The action for a minimally coupled Axion can be written as
\begin{equation}\label{action1}
\textit{A}=\int{\sqrt{-g} d^4x [R + \frac{1}{2}\phi^\mu\phi_\mu - V(\phi) + L_{m}]},
\end{equation}
where $L_{m}$ is the Lagrangian density for ordinary fluid distribution. Energy-momentum contribution from the scalar field $\phi$ therefore is
\begin{equation}\label{minimallyscalar}
T^\phi_{\mu\nu} = \partial_\mu\phi\partial_\nu\phi - g_{\mu\nu}\Bigg[\frac{1}{2}g^{\alpha\beta}\partial_\alpha\phi\partial_\beta\phi - V(\phi)\Bigg]. 
\end{equation}

We assume that the scalar field is spatially homogeneous, i.e., $\phi = \phi(t)$. Therefore we write the field equations as ($8 \pi G = 1$)
\begin{equation} \label{fe1minimal}
3\Bigg(\frac{\dot{a}}{a}\Bigg)^{2} = \rho_{m} + \frac{\dot{\phi}^{2}}{2} + V\left( \phi \right),
\end{equation} 
 
\begin{equation} \label{fe2minimal}
-2\frac{\ddot{a}}{a}-\Bigg(\frac{\dot{a}}{a}\Bigg)^{2} = p_{m} + \frac{\dot{\phi}^{2}}{2}-V\left( \phi \right).
\end{equation}

We also write the scalar evolution equation as
\begin{equation} \label{phiminimal}
\ddot{\phi} + 3\frac{\dot{a}}{a}\dot{\phi} + \frac{dV(\phi)}{d\phi} = 0.  
\end{equation}     
Using the potential in Eq. (\ref{ourpot}) the scalar evolution equation is written as
\begin{equation}\label{minKG}
\ddot{\phi} + 3\frac{\dot{a}}{a}\dot{\phi} + m^{2}\phi - \frac{m^2}{6f^2}\phi^3 = 0.
\end{equation}
From this point on, we call the Klein-Gordon equation for the Axion scalar field evolution as the Axion Evolution Equation. The overhead dot denotes the derivative with respect to $t$. Now Eq. (\ref{minKG}) is clearly a special case of Eq. (\ref{anharmonic}) with $n = 3$ and $f_1 = 3\frac{\dot{a}}{a}$, $f_2 = m^2$ and $f_3 = -\frac{m^2}{6f^2}$. Our progress involves the assumption that we can transform Eq. (\ref{phiminimal}) into an integrable form, writing a transformation like Eq. (\ref{criterion2}). Therefore the criterion in Eq. (\ref{criterion1}) produces an equation governing the allowed dynamics of the radius of two-sphere $a(t)$ and the Eq. (\ref{criterion2}) produces a solution for the scalar field. The equation of the radius of two-sphere is written below

\begin{equation}\label{minimaleq}
\frac{\ddot{a}}{a} + \frac{\dot{a}^2}{a^2} = m^2.
\end{equation}
A simple first integral of this equation can be written as
\begin{equation}
\dot{a}^2 = \frac{m^2 a^2}{2} + \frac{a_0}{a^2},
\end{equation}
where $a_0$ is a non-zero constant of integration. For $\dot{a} < 0$, the above equation is solved to write

\begin{figure}
\begin{center}
\includegraphics[width=0.38\textwidth]{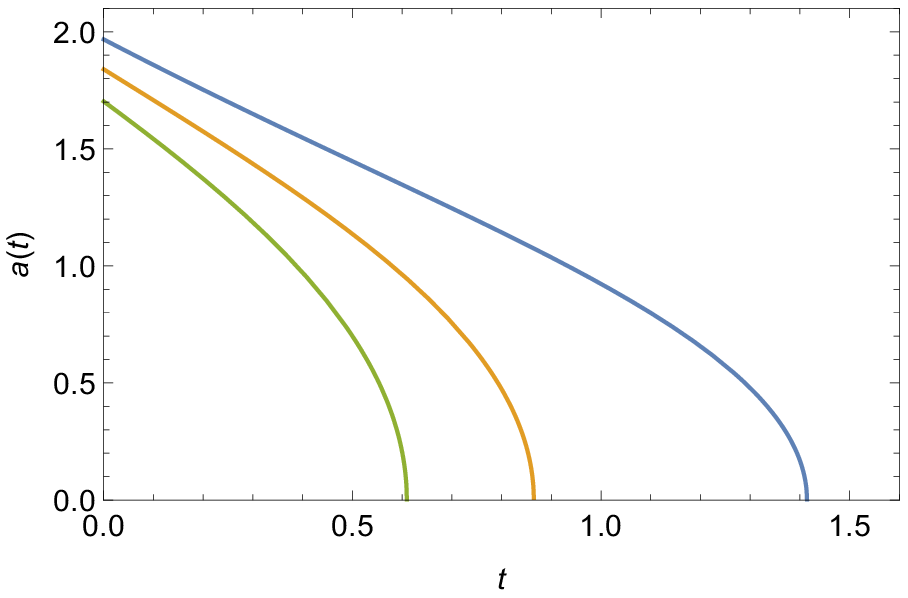}
\includegraphics[width=0.38\textwidth]{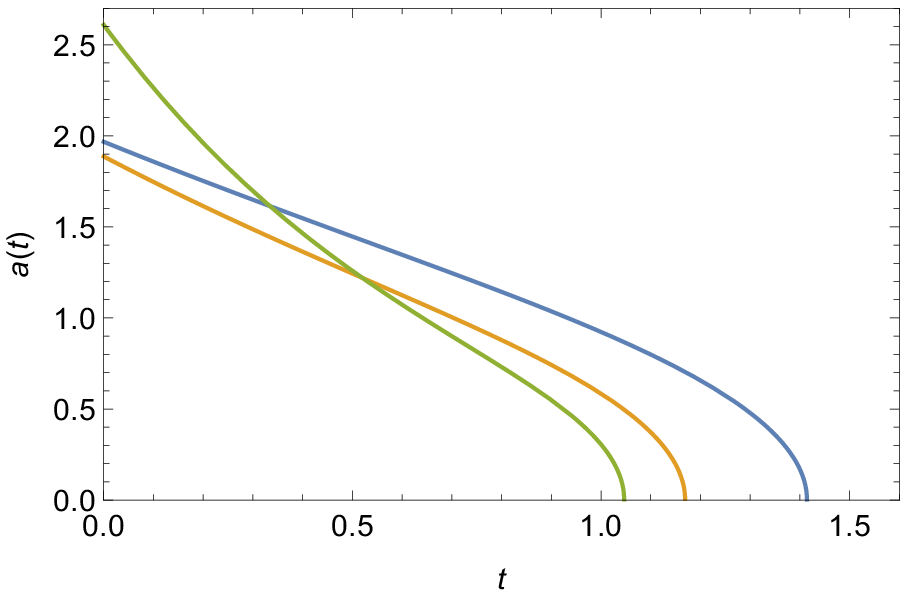}
\caption{Evolution of the two-sphere for a collapsing scalar field. The critical condition for this is $a_0$ is positive. For the graph on top, the choice of $a_{0}$ is varied while $m = \frac{1}{\sqrt{2}}$. For the graph on bottom, the choice of $a_{0}$ is fixed at $a_0 = 1$, while value of $m$ is varied.}
\label{figminimal1}
\end{center}
\end{figure}

\begin{equation}\label{solminimal}
a(t) = \frac{1}{\sqrt{2}m} \Bigg[e^{2m \Big(t_{0} - \frac{t}{\sqrt{2}} \Big)} - 2m^{2}a_{0}e^{-2m \Big(t_{0} - \frac{t}{\sqrt{2}} \Big)}\Bigg]^{\frac{1}{2}}.
\end{equation}

\begin{figure}
\begin{center}
\includegraphics[width=0.38\textwidth]{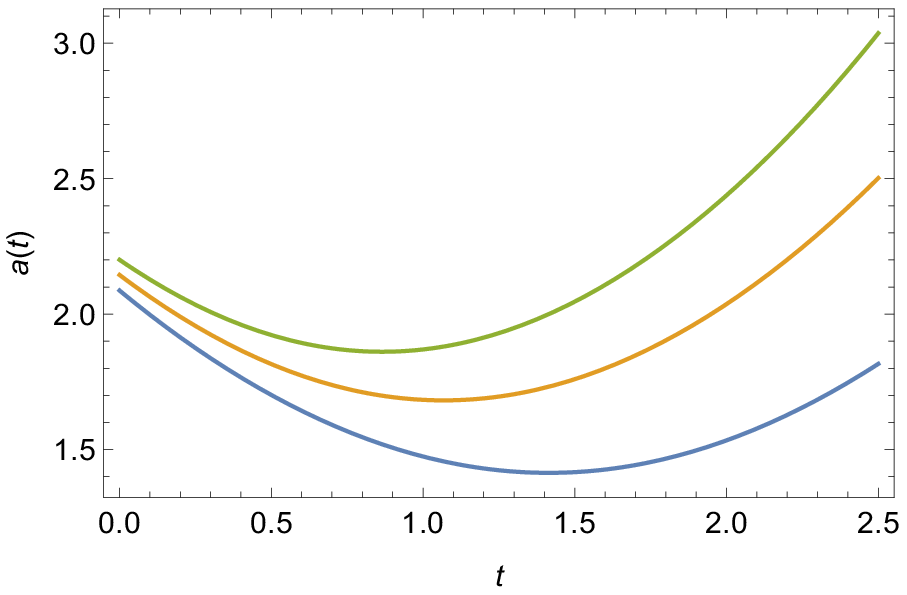}
\includegraphics[width=0.38\textwidth]{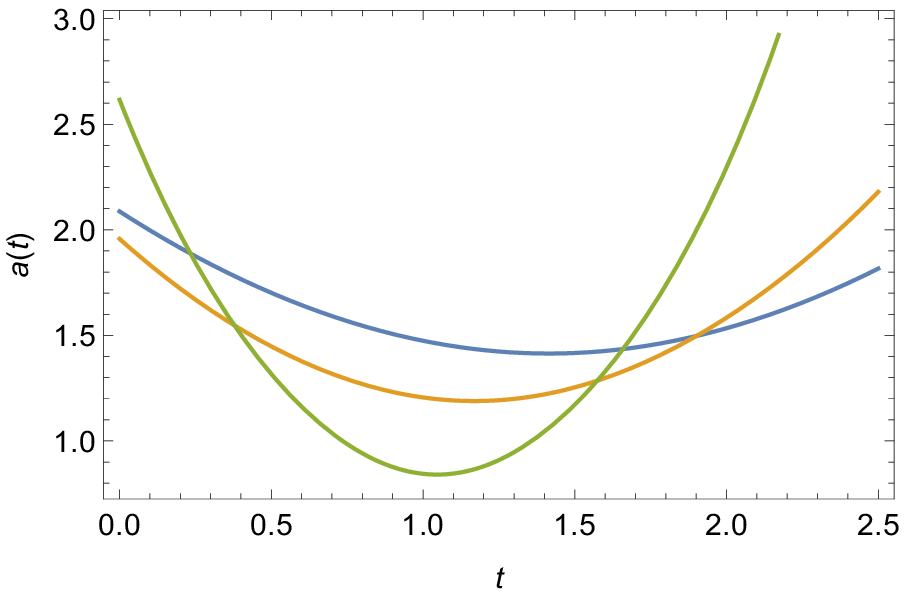}
\caption{Evolution of the two-sphere for a collapsing and bouncing scalar field. The critical condition for this is $a_0$ is negative. For the graph on top, the choice of $a_{0}$ is varied while $m = \frac{1}{\sqrt{2}}$. For the graph on bottom, the choice of $a_{0}$ is fixed at $a_0 = -1$, while value of $m$ is varied.}
\label{figminimal2}
\end{center}
\end{figure}

We plot the evolution of $a(t)$ in Figs. \ref{figminimal1} and \ref{figminimal2} for different initial conditions. Fig. \ref{figminimal1} is the plot of Eq. (\ref{solminimal}) for $a_{0} > 0$. It is clear that the stellar distribution experiences a uniform collapse which speeds up towards the end of it's lifetime and hurries towards a formation of zero proper volume.

The graph on top shows the evolution for a fixed value of $m$, but different choices of $a_0$. On the other hand, the graph on bottom shows the evolution for a fixed value of $a_0 > 0$, but different values of $m$. Fig. \ref{figminimal2} is the plot of Eq. (\ref{solminimal}) for $a_{0} < 0$. In this case, the stellar distribution experiences an initial collapse. Eventually the collapse decelerates and goes through a signature flip of $\dot{a}$ during it's evolution. This indicates a minimum cutoff for the radius after which the collapsing Axion distribution bounces without reaching a zero proper volume. \\

\begin{figure}
\begin{center}
\includegraphics[width=0.37\textwidth]{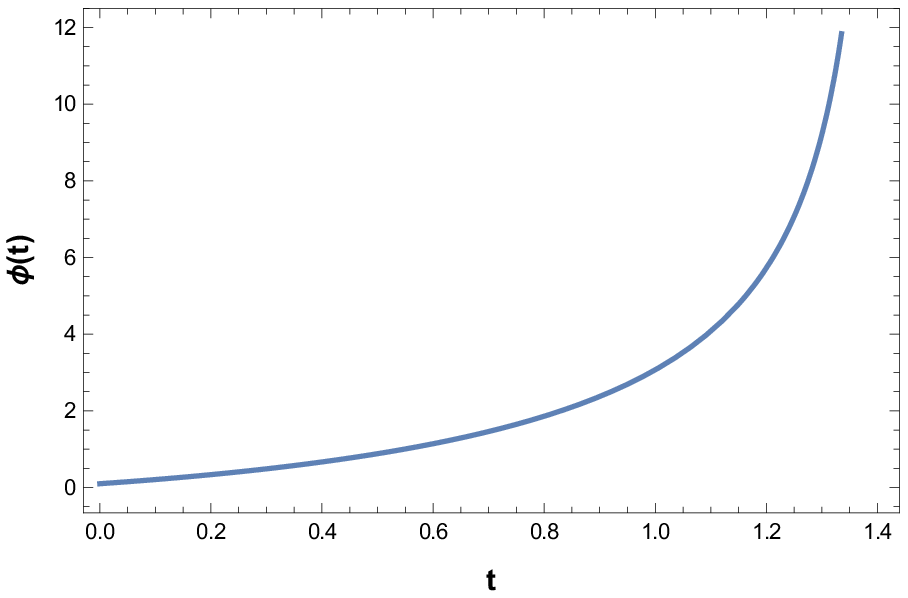}
\includegraphics[width=0.37\textwidth]{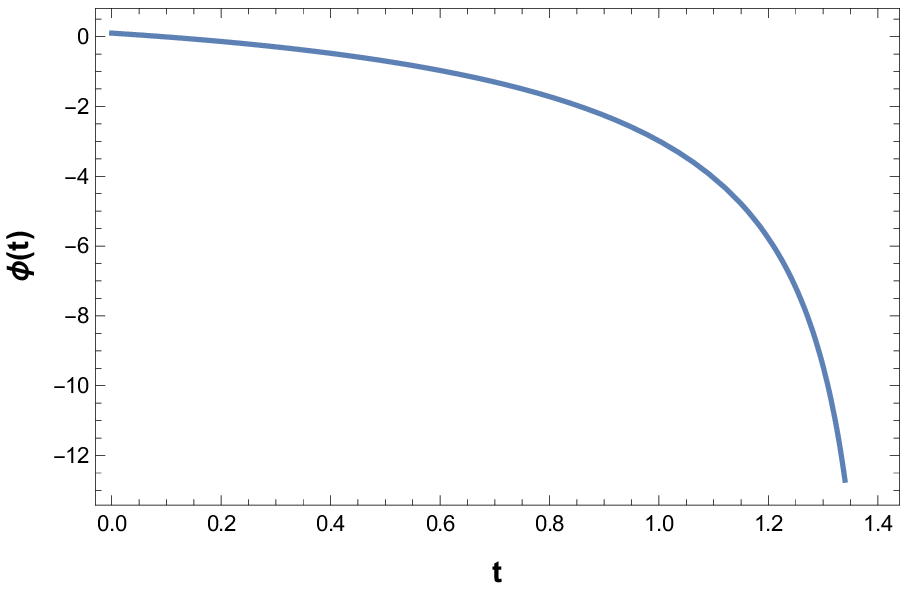}
\caption{Evolution of the scalar field during collapse. The critical condition for this is $a_0$ is positive. For the graph on top, the choice of $a_{0}$ is varied while $m = \frac{1}{\sqrt{2}}$. For the graph on bottom, the choice of $a_{0}$ is fixed at $a_{0} = 1$, while value of $m$ is varied.}
\label{figminimal3}
\end{center}
\end{figure}

\begin{figure}
\begin{center}
\includegraphics[width=0.37\textwidth]{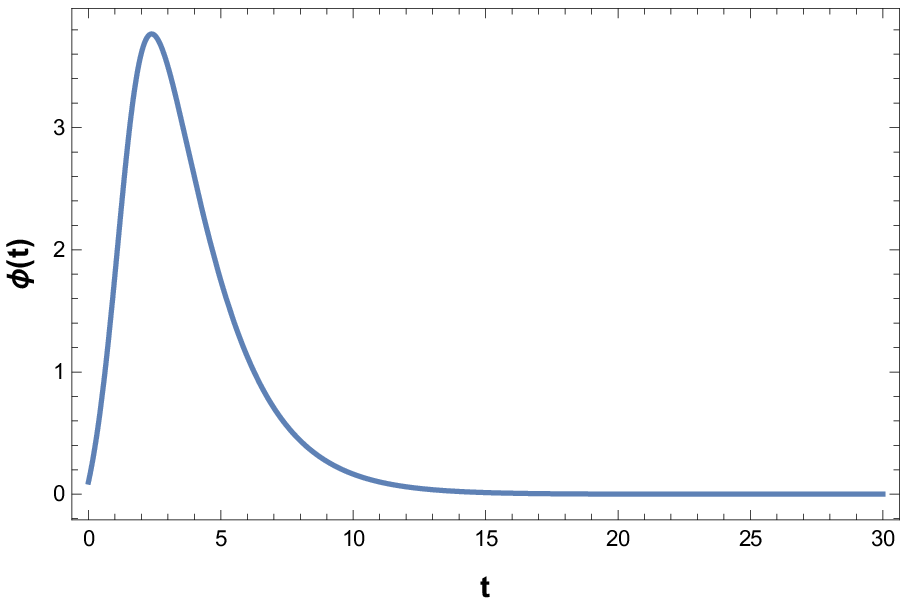}
\includegraphics[width=0.37\textwidth]{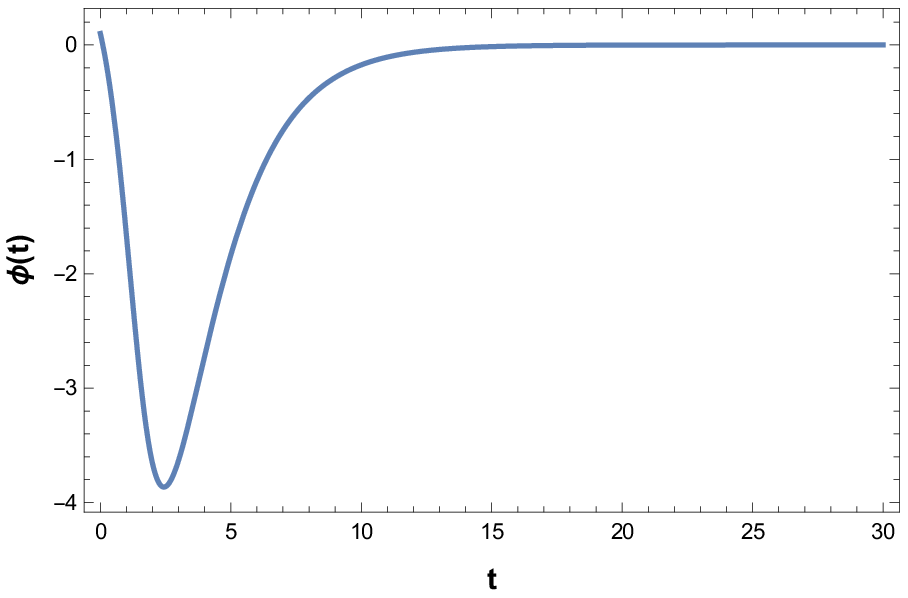}
\caption{Evolution of the scalar field during `collapse and bounce'. The critical condition for this is $a_0$ is negative. For the graph on top, the choice of $a_{0}$ is varied while $m = \frac{1}{\sqrt{2}}$. For the graph on bottom, the choice of $a_{0}$ is fixed at $a_0 = -1$, while value of $m$ is varied.}
\label{figminimal4}
\end{center}
\end{figure}

We study the evolution of the scalar field as a function of time using Eq. (\ref{criterion2}) and transforming the evolution equation Eq. (\ref{phiminimal}). A similar solution can be found by re-inserting the solution Eq. (\ref{solminimal}) in Eq. (\ref{phiminimal}). The exact solution is complicated to write in a closed form. Therefore we resort to numerical study and plot the evolutions in Figs. \ref{figminimal3} and \ref{figminimal4}. The initial conditions chosen for the numerical solutions are the values of $a(t)$ and $\frac{da(t)}{dt}$ for an initial time $t_{i}$. The graphs on top are for initial condition $\frac{da(t)}{dt_{t = t_{i}}} > 0$ and the graphs on bottom are for initial condition $\frac{da(t)}{dt_{t = t_{i}}} < 0$. It is easy to note that for different choice of these initial condition, the evolution of the scalar field only flips in signature, keeping the other qualitative behavior the same. From Fig. \ref{figminimal3} it is clear that for all $a_{0} > 0$, the Axion field diverges at a finite time, just about the time of formation of a zero proper volume. This indicates a formation of a spacetime singularity. On the other hand, if one chooses a negative value of the critical parameter $a_{0}$, the scalar field initially experiences a sharp increasing time evolution. Interestingly, about the time the scale factor changes nature from collapsing to bouncing, the scalar field changes nature as well. It falls off sharply as a function of time, eventually dispersing away to zero value at an asymptotic future (See Fig. \ref{figminimal4}).  \\

The curvature scalars can be written as $R = - 6\left[\frac{\ddot{a}}{a}+\frac{\dot{a}^2}{a^2}\right]$ and $K = 6\left[\frac{\ddot{a}^2}{a^2}+\frac{\dot{a}^4}{a^4}\right]$. Using Eq. (\ref{solminimal}) it is straightforward to check that for $a_{0} > 0$ both Ricci and Kretschmann scalars diverge to infinity when $a(t) \rightarrow 0$. Therefore, if the Axion distribution somehow reaches zero proper volume it definitely ends up in a curvature singularity. Otherwise, it bounces indefinitely after reaching the cutoff radius and ends up radiating away to zero. The distribution of the additional fluid existing alongside the Axion scalar can be studied, courtesy of the field Eqs. (\ref{fe1minimal}) and (\ref{fe2minimal}). We write

\begin{equation}
\rho_{m} = 3\Bigg(\frac{\dot{a}}{a}\Bigg)^{2} - \frac{\dot{\phi}^{2}}{2} -\frac{1}{2}m^{2}\phi^{2} + \frac{m^2}{24f^2}\phi^4,
\end{equation}  
and
\begin{equation}
p_{m} = -2\frac{\ddot{a}}{a}-\Bigg(\frac{\dot{a}}{a}\Bigg)^{2} - \frac{\dot{\phi}^{2}}{2} +\frac{1}{2}m^{2}\phi^{2} - \frac{m^2}{24f^2}\phi^4.
\end{equation}

Therefore the Null Energy Condition (NEC) is
\begin{equation}\label{nec}
\rho_{m} + p_{m} = -2\frac{d}{dt}\Bigg(\frac{\dot{a}}{a}\Bigg) - \dot{\phi}^2.
\end{equation}

It is straightforward that the positivity of the NEC depends on the nature of the first term on the RHS, $-2\frac{d}{dt}\Big(\frac{\dot{a}}{a}\Big)$, since $-\dot{\phi}^2$ is always negative. We assess the nature of $-\frac{d}{dt}\Big(\frac{\dot{a}}{a}\Big)$ from The Raychaudhuri Equation which essentially is a geometric relation to write the dynamics of flows in terms of mean separation between a congruence of curves \cite{rc}. For a system of timelike geodesics whose tangent vector field is given by $u^{\mu}$, the Raychaudhuri equation is
\begin{equation}\label{Raych-time}
 \frac{d\theta}{d\tau} = -\frac{1}{3}\theta^2 + \nabla_{\alpha}a^{\alpha} - \sigma_{\alpha\beta}\sigma^{\alpha\beta} + \omega_{\alpha\beta}\omega^{\alpha\beta} - R_{\alpha\beta}V^\alpha V^\beta,
 \end{equation}

with $\tau$ an affine parameter and $R_{\alpha\beta}$ the Ricci tensor. The acceleration $a_{\alpha}$, expansion $\Theta$ and shear $\sigma_{\alpha\beta}$ of the fluid are given by

\begin{eqnarray}
&& a_{\alpha}=V_{\alpha ;\beta}V^{\beta}, \;\;
\sigma_{\alpha\beta}=V_{(\alpha
;\beta)}+a_{(\alpha}V_{\beta)}-\frac{1}{3}\Theta(g_{\alpha\beta}+V_{\alpha}V
_{\beta}) , \;\; \\&&\nonumber
\Theta={V^{\alpha}}_{;\alpha}.
\end{eqnarray}

The rotation tensor $\omega_{\alpha\beta}$ is defined as

\begin{eqnarray}
&& \omega_{\alpha\beta} = \nabla_{[\alpha}u_{\beta]} - a_{[\alpha}u_{\beta]}, \\&&\nonumber
= \frac{1}{2} \Bigg[\nabla_{\beta}u_{\alpha} - \nabla_{\alpha}u_{\beta} - a_{\alpha}u_{\beta} + a_{\beta}u_{\alpha}\Bigg].
\end{eqnarray}
For a spatially homogeneous spacetime as chosen in the present manuscript, the equation simply leads to
\begin{equation}\label{congruence1}
\frac{d\theta}{d\tau} = 3 \frac{d}{dt}\Bigg(\frac{\dot{a}}{a}\Bigg).
\end{equation}

In general, to realize a collapsing nature $\frac{d\theta}{d\tau}$ must have a negative evolution, eventually reaching $-\infty$ when the singularity forms. Such a criterion is realized only when $\frac{d}{dt}\Big(\frac{\dot{a}}{a}\Big) < 0$. Therefore, the Null Energy Condition in Eq. (\ref{nec}) is satisfied (greater than zero) during the collapse as $\frac{d}{dt}\Big(\frac{\dot{a}}{a}\Big) < 0$ is a necessary condition for the collapse. However, one can not rule out the possibility of $\dot{\phi}$ dominating over $\frac{d}{dt}\Big(\frac{\dot{a}}{a}\Big)$ during the final stages of the collapse since $\phi$ exhibits a sharply increasing profile near the zero proper volume. In cases where the collapse of the Axion field is decelerating, $\frac{d}{dt}\Big(\frac{\dot{a}}{a}\Big)$ remains negative, but the rate of collapse dies down. This can make $\dot{\phi}^2$ dominant enough to violate the Null Energy Condition.  \\

Moreover, the Raychaudhury equation can lead one towards an understanding of the critical behavior of the system. Using Eq. (\ref{solminimal}) (putting $t_{0} = 1$) in Eq. (\ref{congruence1}) one can write 
\begin{equation}
\frac{d\theta}{d\tau} = -\frac{8e^{2m(2+\sqrt{2}t)}m^{4}a_{0}}{(e^{4m} - 2e^{2\sqrt{2}mt}m^{2}a_{0})^2}.
\end{equation}    
It is interesting to note that the value of $m$ makes no contribution in the predictability of the end state, as the signature of $\frac{d\theta}{d\tau}$ depends only on the value of $a_{0}$. Therefore whether the collapsing Axion dark matter keeps on collapsing until a singularity or bounces indefinitely dispersing away all the field distribution to zero value, depends on the parameter $a_0$. This makes a critical parameter of the system, hinting at an underlying critical phenomena.

\section{Non-Minimally Coupled Axion}
In this section we consider an Axion field directly interacting with geometry, apart from having a self-interaction. In that sense it is a bit further generalization from a minimally coupled Axion. We consider an action
\begin{equation}
S =\int d^4x\sqrt{-g}\left[U(\phi)R - \frac{1}{2}g^{\mu\nu}\phi_{,\mu}\phi_{,\nu} + V(\phi) + L_{m}\right].
\label{nonminaction}
\end{equation}

The field equations for a spatially homogeneous metric Eq. (\ref{metric}) are written as
\begin{equation}
\frac{6U\dot{a}^2}{a^2}+\frac{6U'\dot{a}\dot{\phi}}{a}= \rho_{m} + \frac{1}{2}\dot{\phi}^2 + V(\phi),
\label{nonminfrw1}
\end{equation}
and
\begin{equation}
\frac{4U\ddot{a}}{a} + \frac{2U\dot{a}^2}{a^2} + \frac{4U'\dot{a}\dot{\phi}}{a} + 2U''\dot{\phi}^2
+ 2U'\ddot{\phi} = -p_{m} -\frac12\dot{\phi}^2 + V(\phi).
\label{nonminimalfrw2}
\end{equation}

\begin{figure}
\begin{center}
\includegraphics[width=0.38\textwidth]{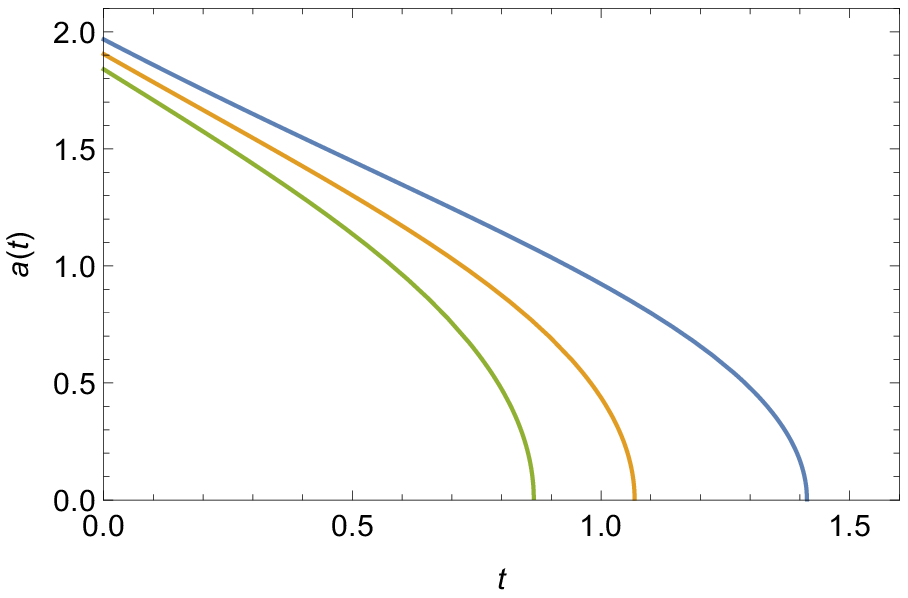}
\includegraphics[width=0.38\textwidth]{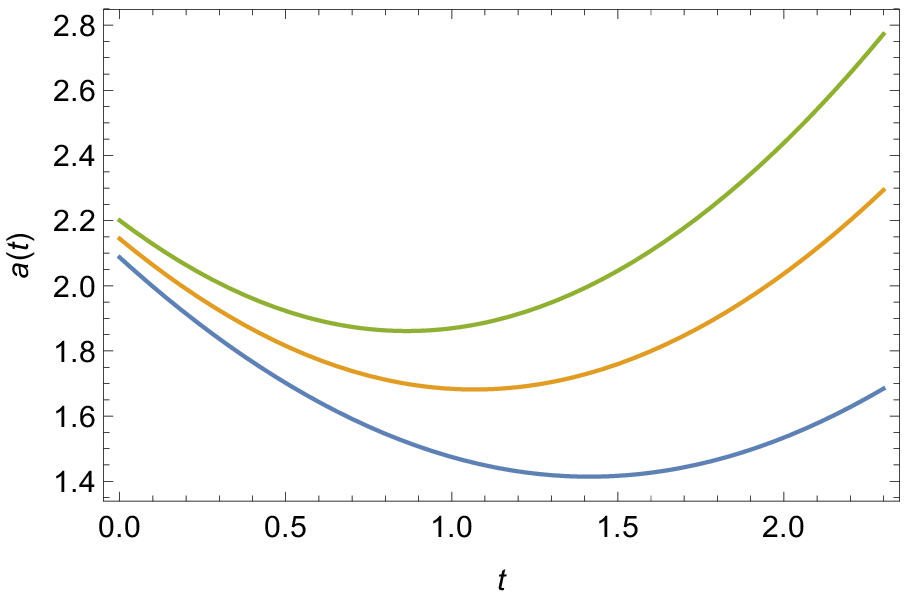}
\caption{Evolution of the two-sphere for $p = \frac{1}{\sqrt{2}}$ and $U_{0} = 2$. For the graph on top, $a_{0} > 0$. For the graph on bottom, $a_{0} < 0$.}
\label{fignonminimal1}
\end{center}
\end{figure}

A variation with respect to the scalar field $\phi$ allows us to write the Axion evolution equation as
\begin{equation}
\ddot{\phi} + 3\frac{\dot{a}}{a}\dot{\phi} - 6U'\left[\frac{\ddot{a}}{a} + \frac{\dot{a}^2}{a^2}\right] + \frac{dV}{d\phi} = 0.
\label{nonminKGini}
\end{equation}

\begin{figure}
\begin{center}
\includegraphics[width=0.37\textwidth]{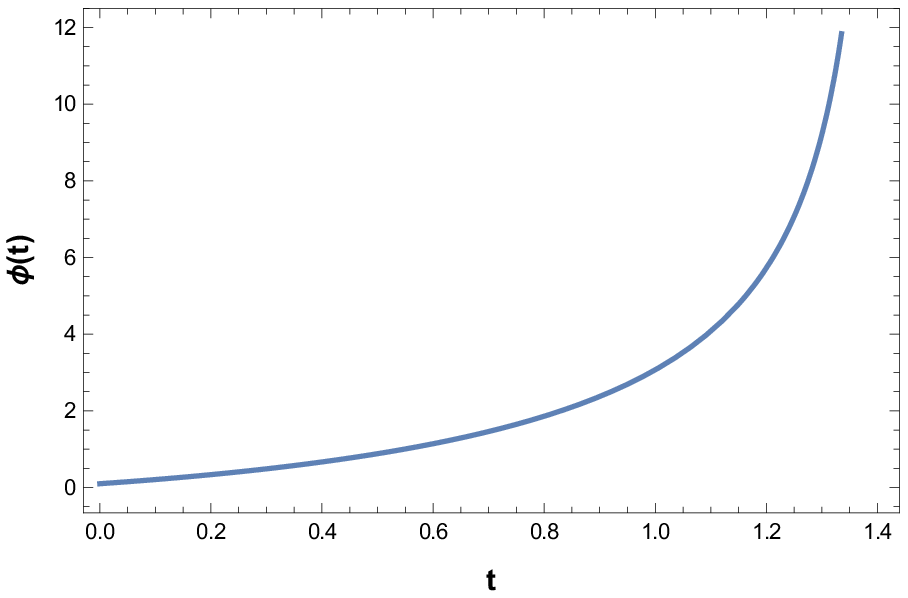}
\includegraphics[width=0.37\textwidth]{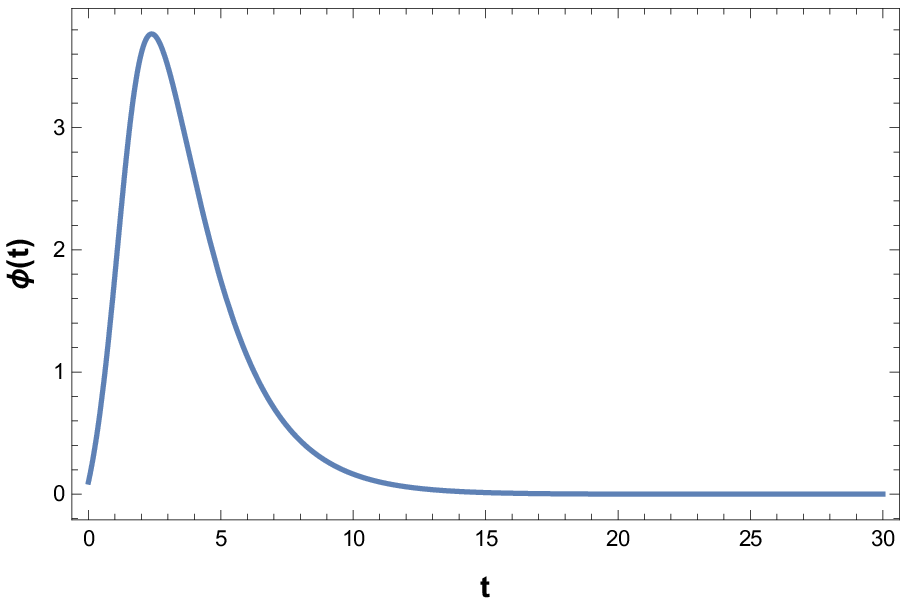}
\caption{Evolution of the scalar field for $p = \frac{1}{\sqrt{2}}$ and $U_{0} = 2$. For the graph on top, $a_{0} > 0$. For the graph on bottom, $a_{0} < 0$.}
\label{fignonminimal2}
\end{center}
\end{figure}

We choose a particular form of the non-minimal coupling function $U(\phi)$ at the outset as
\begin{equation}
U(\phi) = \frac{1}{2}(1 + U_{0}\phi^2).
\end{equation}
The Axion potential is already given by Eq. (\ref{ourpot}). Together with these the Axion evolution equation becomes

\begin{equation}\label{nonminKG}
\ddot{\phi} + 3\frac{\dot{a}}{a}\dot{\phi} + \left[m^{2}-6U_{0} \left(\frac{\ddot{a}}{a} + \frac{\dot{a}^2}{a^2}\right)\right]\phi - \frac{m^2}{6f^2}\phi^3 = 0.
\end{equation}

We follow a similar path to extract a solution, as we did in the previous section. We assume that the Axion evolution equation is integrable and employ the integrability analysis. For brevity we do not repeat the detailed procedure. The integrability criterion produces a second order differential equation of $a(t)$ 
\begin{equation}\label{nonminimaleq}
\frac{\ddot{a}}{a} + \frac{\dot{a}^2}{a^2} = \frac{m^2}{(1+6U_{0})} = p^2.
\end{equation}

This produces a solution for $a(t)$ as
\begin{equation}\label{solnonminimal}
a(t) = \frac{1}{\sqrt{2}p} \Bigg[e^{2p \Big(t_{0} - \frac{t}{\sqrt{2}} \Big)} - 2p^{2}a_{0}e^{-2p \Big(t_{0} - \frac{t}{\sqrt{2}} \Big)}\Bigg]^{\frac{1}{2}}.
\end{equation}

We note that this solution is identical with the minimally coupled Axion case, as can be seen comparing with Eq. (\ref{solminimal}). The only change involves the mass parameter $m$ which is replaced in the non-minimal Axion solution by a rescaled mass parameter, including the non-minimal coupling parameter $U_{0}$, as $m^2 = p^2 (1 + 6U_{0})$. This only scales the time evolution of the stellar evolution and the Axion scalar field while the qualitative behavior remains exactly the same compared to the minimally coupled case. The evolutions are shown in Figs. \ref{fignonminimal1} and \ref{fignonminimal2}. Evolution of the two-sphere in Fig. \ref{fignonminimal1} shows the radius of the two sphere as a function of time for $p = \frac{1}{\sqrt{2}}$ and $U_{0} = 2$. Evolution of the Axion field is shown in Fig. \ref{fignonminimal2} for a similar choice of initial conditions. In both of the graphs, the curve on top is for $a_{0} > 0$ and the curve on the bottom is for $a_{0} < 0$. \\

Using field Eq. (\ref{nonminfrw1}) for density and Eq. (\ref{nonminimalfrw2}) for pressure we can study the nature of the Null Energy Conditions writing
\begin{eqnarray}\nonumber
&&(\rho_{m} + p_{m}) = 4\left(1+U_{0}\phi^{2}\right)\left(\frac{\dot{a}^2}{a^2} - \frac{\ddot{a}}{a}\right) \\&&
+ 2U_{0}\phi \left(\frac{\dot{a}\dot{\phi}}{a} - \ddot{\phi}\right) - \left(1+2U_{0}\right)\dot{\phi}^{2}.
\end{eqnarray}
Although not as simple as the expression for Null Energy Condition for the minimally coupled case, we can assess the nature of the collapsing distribution overall from this, considering some approximations. For a collapsing case, i.e., for all $a_{0} > 0$, as the collapse evolves towards final phase, the radius of two sphere $a(t) \rightarrow 0$ and $\phi, \dot{\phi} \rightarrow \infty$ at a finite time. The rate of collapse increases and this makes $\dot{a}^2$ a sharply increasing and dominating function of time compared to $\ddot{a}$. Therefore, as one approaches the zero proper volume singularity,
\begin{equation}
(\rho_{m} + p_{m}) \sim 4U_{0}\phi^{2}\frac{\dot{a}^2}{a^2} + 2U_{0}\phi\dot{\phi}\frac{\dot{a}}{a} - \left(2U_{0}+1\right)\dot{\phi}^2.
\end{equation}
For collapse $\dot{a} < 0$ and this makes the second and third term on the RHS negative. However, the dominating nature of $\phi^{2}\frac{\dot{a}^2}{a^2}$ is expected to make the RHS positive, satifying the energy condition. \\

However, for a collapse and dispersal scenario, $\dot{a} > 0$ and $\phi, \dot{\phi} \rightarrow 0$ during the dispersal. This allows us to write the energy condition in the form
\begin{equation}
(\rho_{m} + p_{m}) \sim 4\left(\frac{\dot{a}^2}{a^2} - \frac{\ddot{a}}{a}\right) = \frac{16e^{2p(2+\sqrt{2}t)}p^{4}a_0}{(e^{4p}-2e^{2\sqrt{2}pt}p^{2}a_{0})^2},
\end{equation}
where we have chosen $t_{0} = 1$. For a collapse and dispersal scenario, $a_{0} < 0$, and all the other terms on the RHS of the above equation are strictly positive. Therefore, the dispersal of the Axion field and perfect fluid system is usually accompanied by a violation of the NEC. We comment that, since the Axion field distribution almost becomes negligible in this case, the violation of energy condition can be simply an artifact of just the perfect fluid distribution. Usually a dark energy fluid is thought to be violating energy conditions, and from that purview this opens up some intriguing allied questions regarding a clustering dark energy that can coexist with Axion dark matter. Besides, the energy density remains non-negative which means the system is up for any relevant quantization scheme. It is not too unphysical to have a system with violated ebergy conditions as there are examples of non-minimally coupled scalar field systems violateing some or more of the energy conditions \cite{visserbarcelo}.

\section{Axion Interacting with Normal Matter Distribution}
In this section, we work on an idea that under strong gravity, the self-interacting Axion field can interact with ordinary matter, which finds it's motivation from the string inspired dilaton gravity. A similar configuration of scalar field is already popular in a cosmological context, known as Chameleon Fields who can describe a smooth deceleration to acceleration transition of the universe and satisfy the so-called fifth force constraints \cite{khoury}. These models also inspire ideas of interaction between dark energy and dark matter \cite{mota}. A time evolving solution in such a mathematical setup is our primary focus that can reveal the nature of the scalar field during gravitational collapse. We write an action where the Axion scalar field has a non-minimal coupling with Ricci scalar as well as with the ordinary fluid. The action is similar to the Brans-Dicke theories \cite{bd}, a prototype of Scalar-Tensor theories. The generalized action in Jordan frame is 

\begin{eqnarray}\label{1}\nonumber
&& S=\frac{1}{16\pi}\int d^4x \sqrt{-\bar{g}} \{\phi
\bar{R}-\frac{\omega_{bd}}{\phi}\bar{g}^{\mu\nu}\bar{\nabla}_{\mu}\phi
\bar{\nabla}_{\nu}\phi - V(\phi)\\&&
+16\pi f(\phi)L_m\}.
\end{eqnarray} 
where $\bar{R}$ is the Ricci scalar, $\phi$ is a scalar field, $\omega_{bd}$ is the Brans-Dicke parameter, $V(\phi)$ is the self-interaction potential and $f(\phi)$ is an analytic function of the scalar field. We study the system in Einstein frame and employ a conformal transformation to transform the action. We define the transformation as
\begin{equation}\label{a2}
\bar{g}_{\mu\nu}\rightarrow g_{\mu\nu}=\Omega^2 \bar{g}_{\mu\nu},
\end{equation} 
with $\Omega = \sqrt{G \phi}$. We redefine the scalar field as
\begin{equation}\label{a3}
\psi(\phi)=\sqrt{\frac{2\omega_{bd}+3}{16\pi G}} \ln \left(\frac{\phi}{\phi_0} \right).
\end{equation}
$\phi_0\sim G^{-1}$, $\phi > 0$ and $\omega_{bd} > -\frac{3}{2}$ to have real solutions. Therefore in Einstein Frame the action in Eq. (\ref{1}) becomes
\begin{eqnarray}\label{a5}\nonumber
&& S_{EF}= \int d^{4}x \sqrt{-g} \{\frac{R}{16\pi G} -\frac{1}{2}g^{\mu\nu}\nabla_{\mu}\psi\nabla_{\nu}\psi -U(\psi)\\&&
+\exp(-\frac{\sigma\psi}{M_p})~f(\psi) L_{m}\},
\end{eqnarray}
where $\sigma = 8\sqrt{\frac{\pi}{2\omega_{bd} + 3}}$. The self-interaction potential in Einstein frame is therefore written as
\begin{equation}\label{a6}
U(\psi)= V(\phi(\psi))~\exp(-\sigma\psi/M_{p}).
\end{equation}

We now focus entirely on the system in Einstein frame from this point onwards and treat $\psi$ as the Axion scalar field. The self-interaction $U(\psi)$ of $\psi$ is the Axion potential, which we choose to be as in Eq. (\ref{ourpot}). We also define $h(\psi) = e^{-\sigma\psi/M_{p}}f(\psi)$. In a cosmological scenario one has to keep in mind of choosing the parameter $\sigma$ in a manner such that the observational constraints for the $\omega_{bd}$ are satisfied. However for a collapsing solution, the written metric describes an interior geometry of a stellar distribution and therefore such restrictions can be relaxed. Variation of the action (\ref{a5}) with respect to the metric and the Axion field gives the following equations
\begin{equation}\label{2}
G_{\mu\nu} = M_p^{-2} \left(h(\psi)T^{m}_{\mu\nu}+T^{\psi}_{\mu\nu} \right),
\end{equation}
and
\begin{equation}\label{3}
\Box \psi-U'(\psi)=-h'(\psi) L_m.
\end{equation}

The energy momentum contributions of the Axion field and the ordinary fluid are
\begin{equation}\label{4}
T^{\psi}_{\mu\nu} = \left(\nabla_{\mu}\psi\nabla_{\nu}\psi-\frac{1}{2} g_{\mu\nu}\nabla_{\alpha}\psi
\nabla^{\alpha}\psi \right) -U(\psi)g_{\mu\nu},
\end{equation}
and
\begin{equation}\label{5}
T^m_{\mu\nu} = \frac{-2}{\sqrt{-g}}\frac{\delta (\sqrt{-g}L_m)}{\delta g^{\mu\nu}}.
\end{equation}

We write differentiation with respect to $\psi$ as a prime. A calculation of $\nabla^{\mu}G_{\mu\nu}$ using Eq. (\ref{2}) helps us write
\begin{equation}\label{6}
\nabla^{\mu}T^m_{\mu\nu}=(L_m-T^m)\nabla_{\nu}\ln h(\psi),
\end{equation}
where $T^m=g^{\mu\nu}T^m_{\mu\nu}$. This means, covariant derivative of $T^m_{\mu\nu}$ does not vanish, indicating an exchange of energy between matter and the Axion scalar field, depending on the choice of matter Lagrangian density $L_m$. We take a perfect fluid energy-momentum tensor. We also assume that it has an equation of state parameter $w_m$.
\begin{equation}\label{b1}
T^m_{\mu\nu}=(\rho_m+p_m)u_{\mu}u_{\nu}+p_m g_{\mu\nu}.
\end{equation}

The perfect fluid Lagrangian density in the context of GR can be chosen as $L_m = p_m$ or $L_m = -\rho_m$. Whether or not these two choices are equivalent is an intriguing question and has received significant attention in literature over time \cite{lagrangian}. We choose $L_m = -\rho_m$ and write the field equations and the Axion evolution equation using Eqs. (\ref{2}) and (\ref{3}) for metric Eq. (\ref{metric}) 

\begin{equation}\label{a11}
3\left(\frac{\dot{a}}{a}\right)^2=M_p^{-2}(h(\psi)\rho_{m}+\rho_{\psi}),
\end{equation}

\begin{equation}\label{a12}
2{\left(\frac{\ddot{a}}{a}\right)} + \left(\frac{\dot{a}}{a}\right)^2=-M_p^{-2}~(h(\psi)p_{m}+p_{\psi}),
\end{equation}
and
\begin{equation}\label{a13}
\ddot{\psi}+3\left(\frac{\dot{a}}{a}\right)\dot{\psi}+\frac{dU(\psi)}{d\psi} + \rho_{m}h'(\psi) = 0.
\end{equation}

To realize the underlying physics we also write the matter conservation equation as

\begin{equation}\label{a14}
\dot{\rho}_{m}+3H(\omega_m+1)\rho_m=-(1-2\omega_m)\dot{\psi}\frac{h'}{h}\rho_m.
\end{equation}

\begin{figure}
\begin{center}
\includegraphics[width=0.38\textwidth]{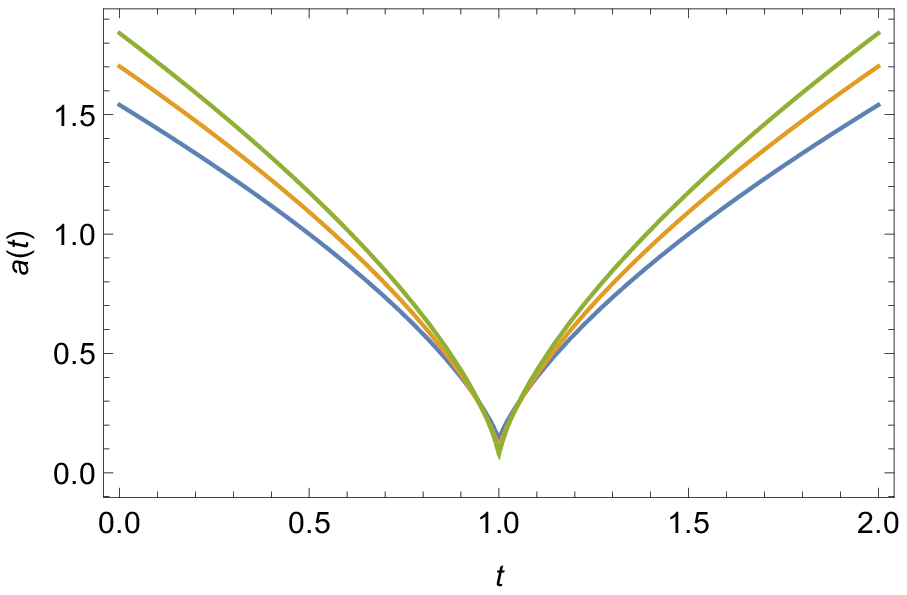}
\includegraphics[width=0.38\textwidth]{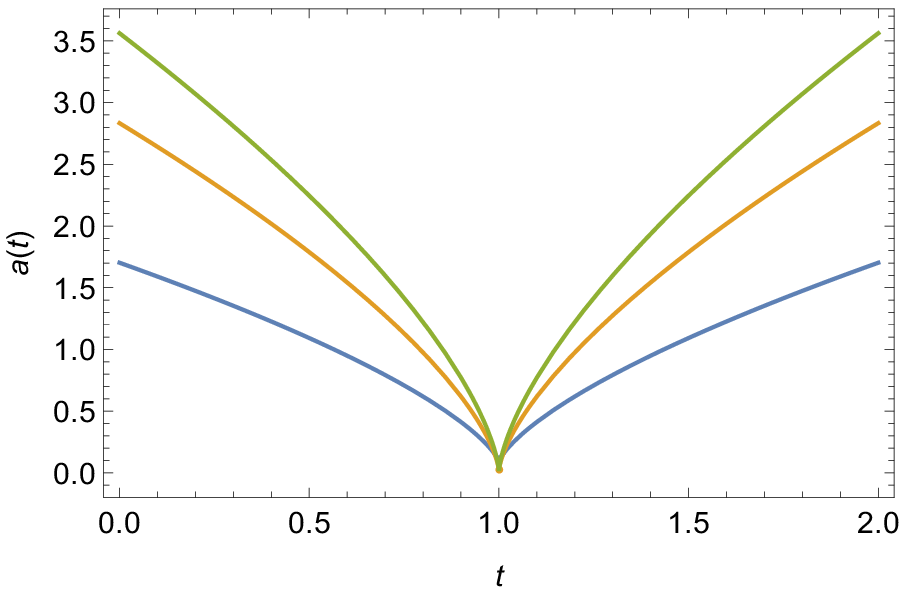}
\caption{Time evolution of the two-sphere for scalar-matter interaction. For the graph on top, the choice of $\rho_{0}$ is varied while other parameters are kept fixed. For the graph on bottom, the choice of $\psi_{0}$ is varied while other parameters ar kept fixed.}
\label{figcham1}
\end{center}
\end{figure}

We integrate Eq. (\ref{a14}) once and write,

\begin{equation}\label{a166}
\rho_m = \rho_{0} a^{-3(\omega_m+1) + \epsilon}.
\end{equation}

The function $\epsilon$ is defined as 

\begin{eqnarray}\label{a16}
&& \epsilon = \frac{(2\omega_m-1)}{M_p}\frac{\int \beta d\psi}{\ln a}, \\&&
\frac{h'(\psi)}{h(\psi)} \equiv \frac{\beta(\psi)}{M_p}.
\end{eqnarray}
$\rho_{0}$ is an integration constant. Interestingly, the Eqs. (\ref{a166}) and (\ref{a16}) indicate that the energy density evolves in a non-trivial manner as compared to GR, due to scalar interaction. For $\epsilon < 0$, we can say that there is matter creation and energy is fed into matter. Similarly, for $\epsilon > 0$ it can be thought that matter is annihilated. There is an outwards energy transfer from the matter. 

\begin{figure}
\begin{center}
\includegraphics[width=0.38\textwidth]{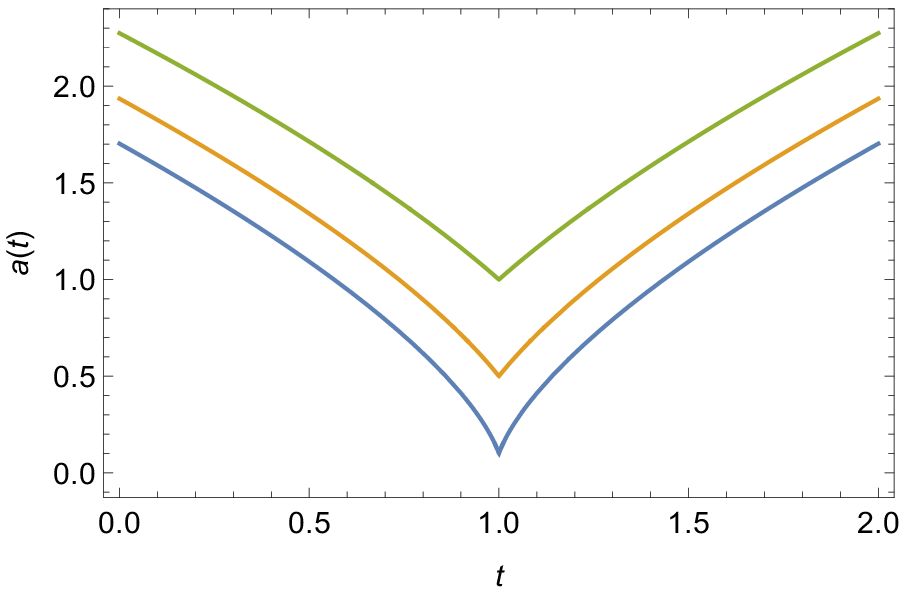}
\includegraphics[width=0.38\textwidth]{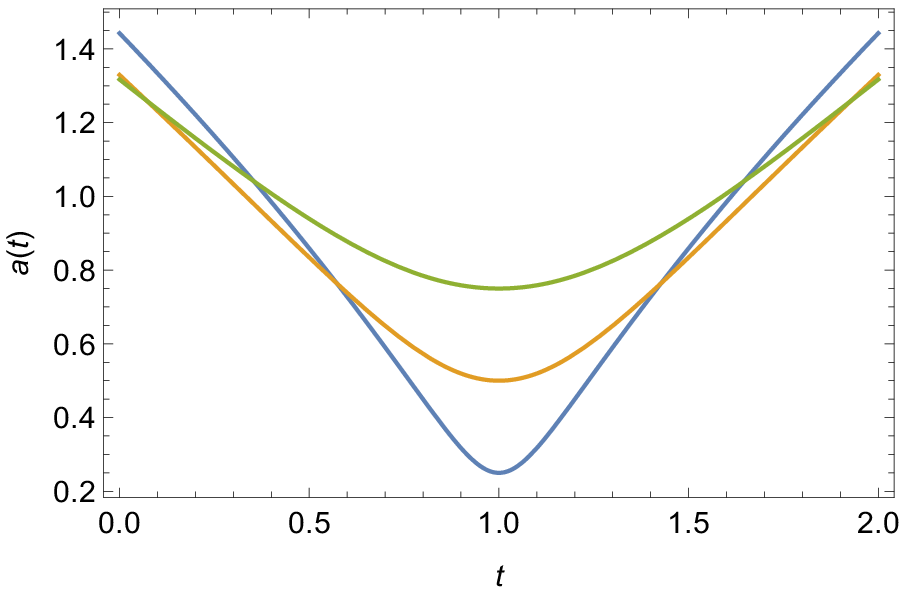}
\caption{Time evolution of the two-sphere for scalar-matter interaction. For the graph on top, the choice of $a_{0} > 0$ is varied while other parameters are kept fixed. For the graph on bottom, $a_{0} < 0$ is varied while other parameters ar kept fixed.}
\label{figcham2}
\end{center}
\end{figure}

The Axion evolution equation Eq. (\ref{a13}) has a contribution from the non-trivial scalar-matter interaction, which is evident from the terms $\rho_{m}$ and $h'(\psi)$ on the LHS. For simplicity we assume the ordinary matter distribution inside the collapsing distribution to be dust, i.e., $w_m = 0$. A careful look into Eq. (\ref{a16}) reveals that, during the final stages of a collapse, $a \rightarrow 0$, implies $\frac{\int \beta d\psi}{\ln a} \rightarrow 0$. This implies that $\rho_m \propto \frac{1}{a^3}$ during the final phases of the collapse. Using this simplification restricts the allowed time evolution, nevertheless, it allows us to solve the set of equations to extract a solution of significant importance. We assume the Axion interaction term to be $h(\psi) = \frac{\psi_{0}\psi^{2}}{2}$ and assume that the Axion evolution equation is integrable. Moreover, we write the Axion self-interaction potential as in Eq. (\ref{ourpot}). With all these, the Axion evolution equation finally becomes
\begin{equation}\label{psichameleon}
\ddot{\psi}+3\left(\frac{\dot{a}}{a}\right)\dot{\psi} + \left(m^2 + \frac{\rho_{0}\psi_{0}}{a^3}\right)\psi - \frac{m^2}{6f^2}\psi^{3} = 0.
\end{equation}

The integrability criterion produces a second order differential equation of $a(t)$ as follows
\begin{equation}\label{chamminimaleq}
\frac{\ddot{a}}{a} + \frac{\dot{a}^2}{a^2} = m^{2} + \frac{\rho_{0}\psi_{0}}{a^{3}}.
\end{equation}
The first integral from the equation is written as
\begin{equation}\label{firstintcham}
\dot{a}^2 = \frac{m^{2}a^{2}}{2} + \frac{2\rho_{0}\psi_{0}}{a} + \frac{a_{0}}{a^2}.
\end{equation}

An exact solution of this equation can not be written in closed form, however, we note that during the final stages of the collapse, $a(t) \rightarrow 0$ and thus $\frac{2\rho_{0}\psi_{0}}{a} + \frac{a_{0}}{a^2}$ clearly dominates over $\frac{m^{2}a^{2}}{2}$. With this in mind, we can ignore the first term on the RHS and solve the equation to write

\begin{widetext}
\begin{eqnarray}\nonumber\nonumber\label{exactchameleon}
&& a(t) = \frac{a_{0}}{2\rho_{0}\psi_{0}} + \frac{3a_{0}^{2}}{2^{2/3}\rho_{0}^{2}\psi_{0}^{2}} \Bigg[\Bigg(-1944 c_1 \rho_{0} \psi_{0} t + 972 c_{1}^2 \rho_{0} \psi_{0} - \frac{54 a_{0}^{3}}{\rho_{0}^{3} \psi_{0}^{3}} + 972 \rho_{0} \psi_{0} t^{2} \Bigg) + \Bigg\lbrace -\frac{2916 a_{0}^6}{\rho_{0}^6 \psi_{0}^6} + \Bigg(-1944 c_{1} \rho_{0} \psi_{0} t \\&&\nonumber
+ 972 c_{1}^{2} \rho_{0} \psi_{0} - \frac{54 a_{0}^{3}}{\rho_{0}^3  \psi_{0}^{3}} + 972 \rho_{0} \psi_{0} t^2 \Bigg)^{2}\Bigg\rbrace^{1/2} \Bigg]^{-1/3} + \frac{1}{6 \times 2^{1/3}} \Bigg[\Bigg(-1944 c_1 \rho_{0} \psi_{0} t + 972 c_{1}^2 \rho_{0} \psi_{0} - \frac{54 a_{0}^{3}}{\rho_{0}^{3} \psi_{0}^{3}} + 972 \rho_{0} \psi_{0} t^{2} \Bigg) \\&&
+ \Bigg\lbrace -\frac{2916 a_{0}^6}{\rho_{0}^6 \psi_{0}^6} + \Bigg(-1944 c_{1} \rho_{0} \psi_{0} t + 972 c_{1}^{2} \rho_{0} \psi_{0} - \frac{54 a_{0}^{3}}{\rho_{0}^3  \psi_{0}^{3}} + 972 \rho_{0} \psi_{0} t^2 \Bigg)^{2}\Bigg\rbrace^{1/2} \Bigg]^{1/3}.
\end{eqnarray}
\end{widetext}

\begin{figure}
\begin{center}
\includegraphics[width=0.38\textwidth]{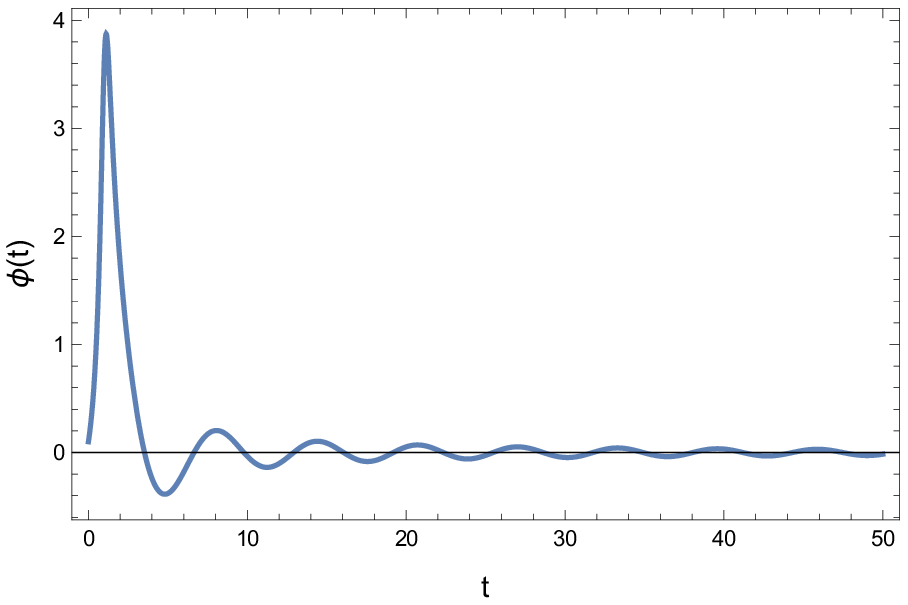}
\includegraphics[width=0.38\textwidth]{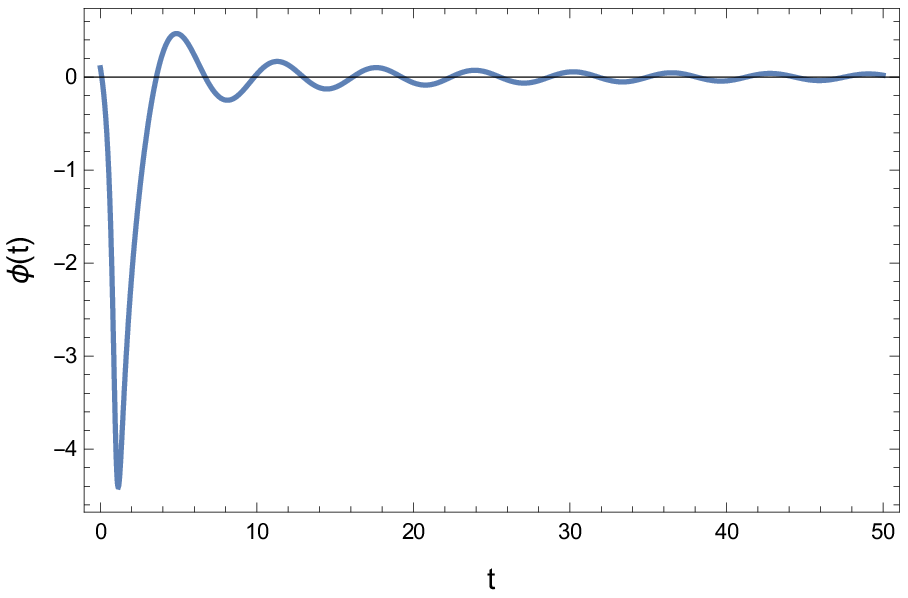}
\caption{Time evolution of the Axion field for all $a_{0} < 0$. For the graph on top, the initial condition $\frac{da}{dt}_{t=t_i} > 0$ while all other parameters are kept fixed. For the graph on bottom, $\frac{da}{dt}_{t=t_i} < 0$ while all other parameters are kept fixed.}
\label{figcham3}
\end{center}
\end{figure}

\begin{figure}
\begin{center}
\includegraphics[width=0.38\textwidth]{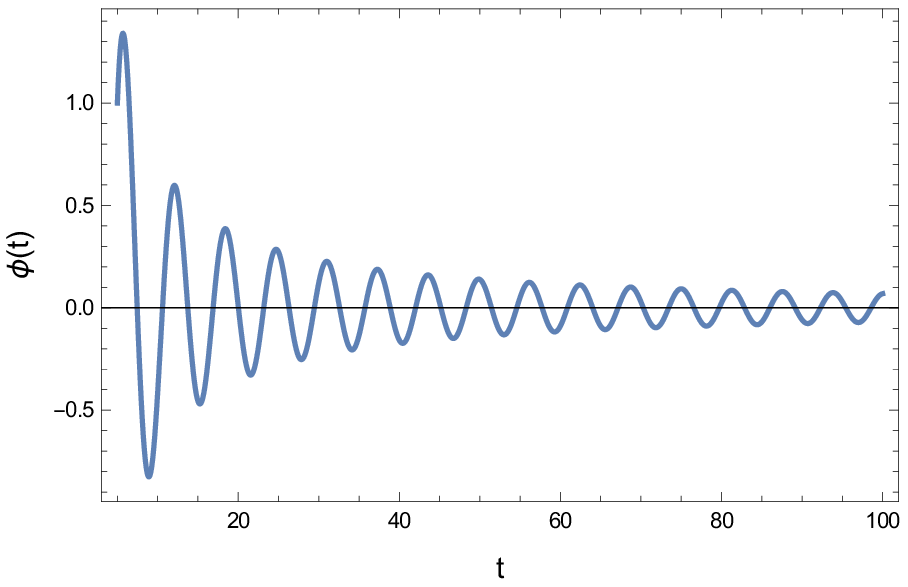}
\includegraphics[width=0.38\textwidth]{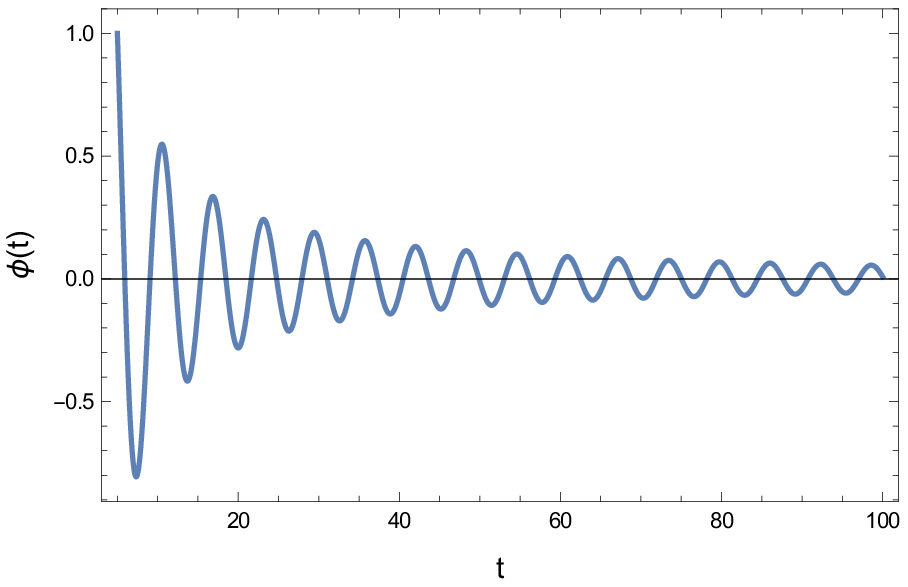}
\caption{Time evolution of the Axion field for all $a_{0} > 0$. For the graph on top, the initial condition $\frac{da}{dt}_{t=t_i} > 0$ while other parameters are kept fixed. For the graph on bottom, $\frac{da}{dt}_{t=t_i} < 0$ while other parameters are kept fixed.}
\label{figcham4}
\end{center}
\end{figure}

The evolution of $a(t)$ is shown in Figs. \ref{figcham1} and \ref{figcham2} as a function of time. In Fig. \ref{figcham1}, the evolution is shown for different choices of $\rho_{0}$ (top graph) and $\psi_{0}$ (bottom graph) while all other parameters ar kept fixed. On the other hand, in Fig. \ref{figcham2}, the plot is for different choices of the parameter $a_{0}$. The graph on top shows the evolution for all $a_{0} > 0$ and the graph on bottom shows the evolution for $a_{0} < 0$ while all the other parameters ar kept fixed. It is interesting to see that the radius of two-sphere $a(t)$ never reaches a zero proper volume for any combination of choices of initial conditions. The collapse always halts after a finite time, at a non-zero minimum cutoff value, after which it bounces. \\

We study the evolution of the Axion scalar as a function of time using Eq. (\ref{criterion2}) and the evolution equation Eq. (\ref{psichameleon}). The same solution is found by re-inserting the solution in Eq. (\ref{exactchameleon}) in Eq. (\ref{psichameleon}) for a consistency check. The solution is complicated to write in a closed form. Therefore we resort to numerical study and plot the evolutions in Figs. \ref{figcham3} and \ref{figcham4}. The initial conditions chosen for the numerical solutions are the values of $a(t)$ and $\frac{da(t)}{dt}$ for an initial time $t_{i}$. The graphs on top are for initial condition $\frac{da(t)}{dt_{t = t_{i}}} > 0$ and the graphs on bottom are for initial condition $\frac{da(t)}{dt_{t = t_{i}}} < 0$. It is easy to note that for different choice of these initial condition, the evolution of the scalar field only flips in signature, keeping the qualitative behavior same. From Fig. \ref{figcham3} and \ref{figcham4} we note that the Axion field initially goes through a sharp increasing/decreasing time evolution, depending on the choice of $\frac{da(t)}{dt_{t = t_{i}}}$. Just about the time the scale factor changes nature from collapsing to bouncing, the magnitude of the scalar field falls off sharply as a function of time and the field starts showing an oscillatory profile. The amplitude of the oscillations eventually dies down and the Axion field disperses away to zero value at an asymptotic future. \\

Using the field Eqs. (\ref{a11}) and (\ref{a12}) we write the NEC as

Therefore we can write the NEC as
\begin{equation}\label{neccham}
\rho_{m} + p_{m} = - \frac{2\frac{d}{dt}\left(\frac{\dot{a}}{a}\right) + \dot{\psi}^2}{h(\psi)} = - 2\frac{2\frac{d}{dt}\left(\frac{\dot{a}}{a}\right) + \dot{\psi}^2}{\psi_{0} \psi^{2}}.
\end{equation}

The Axion field $\psi$ rapidly decays throughout the collapse and therefore the role of $\dot{\psi}^2$ eventually becomes less dominant compared to the first term of the numerator. We note that the choice of $\psi_{0}$, indicates the choice of the Axion-ordinary matter interaction, and plays a crucial part in determining the positive or negative signature of the LHS, and therefore, the validity of the NEC.

\section{Matching with an exterior Vaidya Spacetime}
A collapsing stellar distribution must be surrounded by a suitably chosen exterior spacetime geometry. It is not unphysical to think that the exterior is almost vacuum. For a spherically symmetric case, a Schwarzschild solution is the most popular choice to fit in as the exterior and the metric and the extrinsic curvature is matched across the boundary hypersurface following the standard Israel-Durmois methodology \cite{israel}. With a scalar field however, this is less straightforward, as a Schwarzschild solution supports no scalar field \cite{sc3}. We find it more reasonable to match the interior solutions in the present case with an exterior Vaidya metric across a boundary hypersurface $\Sigma$ \cite{noscalar}. We write the interior metric as
\begin{equation}\label{interior}
d{s_-}^2=dt^2-a(t)^2dr^2-r^2 a(t)^2d{\Omega}^2,
\end{equation}
The exterior Vaidya metric is
\begin{equation}\label{exterior}
d{s_+}^2 = \left(1-\frac{2 M(r_v,v)}{r_v} \right)dv^2 + 2dvdr_v - {r_v}^2d{\Omega}^2.
\end{equation}

In all the three examples discussed in the present manuscript, the basic metric structure of the interior follows the structure of Eq. (\ref{interior}), only the time dependence of $a(t)$ changes. Therefore we deal with the matching condition for a general $a(t)$. Continuity of metric or the first fundamental form gives
\begin{equation}\label{cond1}
\Big(\frac{dv}{dt} \Big)_{\Sigma}=\frac{1}{\sqrt{1-\frac{2M(r_v,v)}{r_v}+\frac{2dr_v}{dv}}}
\end{equation}
and
\begin{equation}\label{cond2}
(r_v)_{\Sigma} = r a(t).
\end{equation}

Eq. (\ref{cond2}) is the first matching condition. Continuity of the extrinsic curvature on the boundary hypersurface gives
\begin{equation}\label{cond3}
\Big(r a(t) \Big)_{\Sigma} = r_v\left(\frac{1-\frac{2M(r_v,v)}{r_v}+\frac{dr_v}{dv}}{\sqrt{1-\frac{2M(r_v,v)}{r_v}+\frac{2dr_v}{dv}}}\right).
\end{equation}

We write, combining Eqs. (\ref{cond1}), (\ref{cond2}) and (\ref{cond3}) 
\begin{equation}\label{dvdt2}
\left(\frac{dv}{dt} \right)_{\Sigma} = \frac{3r a(t)^2 - r^2}{3(ra(t)^2 - 2Ma(t))}.
\end{equation}

Eq. (\ref{dvdt2}) is the second matching condition. We write from Eq. (\ref{cond3})
\begin{equation}\label{M}
M_{\Sigma} = \frac{1}{4} \Bigg[ra(t) + \frac{r^3}{9 a(t)^3} + \sqrt{\frac{1}{r a(t)} + \frac{r^3}{81 a(t)^9} - \frac{2r}{9 a(t)^5}} \Bigg].
\end{equation}

Continuity of extrinsic curvature leads to the derivative of $M(v,r_v)$
\begin{equation}\label{dM}
M{(r_v,v)}_{,r_v}=\frac{M}{r a(t)} - \frac{2r^2}{9 a(t)^{4}}.
\end{equation}
Eqs. (\ref{M}) and (\ref{dM}) are the third and fourth matching conditions. The three different cases presented in the manuscript are three specific cases of this general conditions.

\section{Conclusion}
The manuscript deals with the subject of stellar collapse under the scope of classical gravity. The topic has received rigorous and recurring attempts over the years for possible explanations of some of the inconclusive questions of gravitational physics. For example, the correct time evolution of a collapsing stellar distribution or the predictability of a collapse end-state has baffled physicists. The main difficulty remains the high non-linearity of the field equations from which an extraction of time-dependent solutions is never a cakewalk. Therefore, we find it sensible to try and understand the dynamics in small and gradual steps, using relatively simpler setups that can describe the underlying physics. We consider the gravitational collapse of an interacting time-dependent scalar field distribution alongwith a perfect fluid and gradually generalize and include non-trivial interactions of the scalar with geometry and even ordinary matter. \\

To read through the non-linearity, we point transform the Klein-Gordon type equation for the scalar field evolution into a totally integrable form using the Euler-Duarte-Moreira analysis for anharmonic oscillator equation systems \cite{euler1, euler2, harko}. Application of this method leads us to some new interesting solutions of the radius of the two-sphere for the metric tensor and the scalar field. We deal with a simple, spatially homogeneous Oppenheimer-Snyder type metric. For a minimally coupled self-interacting scalar field, we find an equally likely probabilty that the collapse can form a singular end-state or a bounce, depending on an initial collapsing parameter. The parameter comes out as a constant of integration and can be related to the initial value of the radius of two-sphere or the initial volume of the stellar distribution (the parameter $a_0$). We predict this parameter to be a critical parameter and catch a hint of an unexplored critical phenomena of the system. We predict a similar end result for a second example when the scalar field interacts with the geometry through a non-minimal coupling, apart from it's self-interaction. In a third example, the scalar field interacts with geometry as well as the normal matter distribution through a non-minimal coupling function. For no functional form of the self-interaction potential or the non-minimal interaction functions or for no value of the parameters of the metric coefficients, any zero proper volume singularity is reached. This serves as an example of a rare case of non-singular gravitational collapse, where the collapse can only go unhindered until a minimum accessible volume of the spherical geometry. After this point, the evolution changes nature and bounces indefinitely. The scalar field has a periodic evolution as a function of time with changing amplitude and frequency. Eventually the scalar field disperses away all of it's field strength and asymptotically reaches zero.      \\ 

We note that, if and when a singular end state develops, the visibility of the same is dependent on the existence of nonspacelike trajectories coming out of the singular epoch and reaching a faraway observer. In the present case, for all singularity reaching solutions, the time at which a zero proper volume is reached is independent of the radial coordinate. This indicates that all the collapsing shells consisting of the scalar field and perfect fluid fall onto the singularity simulteneously. This can only form a covered singularity or a Black Hole \cite{maartens}. \\

The investigation stems from an idea to study the evolution of the QCD inspired `Axion' field under strong gravity. The Axion field can actually serve as a very good fit for the cosmological `dark' matter distribution of the universe, the driver of cosmic deceleration. The scalar field and the self-interaction potential used throughout this manuscript describes the evolution of Axion dark matter in spherical symmetry. Our discussion effectively leads one to realize the possibility of Axion dark matter to collapse and form singularities, gravitationally bound objects or explode and disperse away into zero field strength, under strong gravity. We find two possible end scenarios. (a) `Axion Black Holes', where a zero proper volume singularity forms in the end. Such a singularity is always accompanied by the formation of an apparent horizon as it is supposed to be for a spatially homogeneous geometry. (b) `A Collapse and Dispersal of Axion field', where the collapse ends at a non zero minimum radius of two-sphere and the stellar distribution begins to bounce indefinitely, dispersing away all the Axion scalar field strength. The second case can be roughly compared with the phenomena of a stellar distribution radiating energy away from itself through bursts of Axions, mimicking the `Bosenova phenomena' of cold-atom physics \cite{bosenova}. The accompanying perfect fluid distribution in the models can somewaht be connected with a Dark Energy fluid as we study it's distribution and evolution during the collapse. We study the violation of energy condition for this fluid description and show that even for perfectly reasonable physical conditions, it is natural for this fluid to violate atleast the Null Energy Conditions. This serves as a hint towards a couple scenarios. (a) A collapsing fluid violating energy conditions under any circumstances is suggestive of a possible clustering nature of Dark Energy. (b) A stellar distribution involving a Scalar Dark Matter and a fluid Dark Energy combination under strong gravity can lead one to understand the nature of gravitational interaction in a cosmic era where both of these dark components were dominant and interacting with one another. 

\section{Acknowledgements}
The author would like to thank Professor David Fonseca Mota for useful comments and suggestions.

\end{document}